\documentclass[cernpreprint,atlasdraft=false,texlive=2017,USenglish,texmf]{atlasdoc}
 
\usepackage[subfigure=true, biblatex=true, backend=bibtex]{atlaspackage}
 
\usepackage{atlasbiblatex}
 
\usepackage{atlasphysics}
 
\graphicspath{{logos/}{figures/}}
 
\usepackage{ANA-HDBS-2018-37-PAPER-defs}
 
\usepackage{multirow}
 
\AtlasTitle{Search for Higgs Boson Decays into a $Z$~Boson and a Light Hadronically Decaying Resonance Using 13~TeV $pp$ Collision Data from the ATLAS Detector}
 
\AtlasAbstract{
A search for Higgs boson decays into a $Z$~boson and a light resonance in two-lepton plus jet events is performed, using a $pp$ collision dataset with an integrated luminosity of $139~\text{fb}^{-1}$ collected at $\sqrt{s}=13\ \text{TeV}$ by the ATLAS experiment at the CERN LHC. The resonance considered is a light boson with a mass below 4~GeV from a possible extended scalar sector, or a charmonium state. Multivariate discriminants are used for the event selection and for evaluating the mass of the light resonance. No excess of events above the expected background is found. Observed (expected) 95\% confidence-level upper limits are set on the Higgs boson production cross section times branching fraction to a $Z$~boson and the signal resonance, with values in the range $17\text{--}340~\text{pb}$ ($16^{+6}_{-5}\text{--}320^{+130}_{-90}~\text{pb}$) for the different light spin-0 boson mass and branching fraction hypotheses, and with values of $110$ and $100~\text{pb}$ ($100^{+40}_{-30}$ and $100^{+40}_{-30}~\text{pb}$) for the $\eta_c$ and $J/\psi$ hypotheses, respectively.
}
 
\author{ATLAS Collaboration}
 
\PreprintIdNumber{CERN-EP-2020-033}
 
\AtlasJournalRef{Phys. Rev. Lett. 125 (2020) 221802}
\AtlasDOI{10.1103/PhysRevLett.125.221802}
 
\hypersetup{pdftitle={ATLAS document},pdfauthor={The ATLAS Collaboration}}
 
\begin{document}
 
\maketitle
 
The structure of the standard model (SM) scalar sector is the subject of intense scrutiny by the ATLAS~\cite{PERF-2007-01} and CMS~\cite{CMS-TDR-08-001} Collaborations at the CERN Large Hadron Collider (LHC)~\cite{Evans:2008zzb}. At the current level of precision, all of the measured properties of the Higgs boson ($H$)~\cite{HIGG-2012-27,CMS-HIG-12-028} are found to be consistent with their SM predictions~\cite{HIGG-2014-14,HIGG-2013-17-witherratum,CMS-HIG-14-032,HIGG-2017-06,HIGG-2015-07}, and no additional Higgs boson has been observed to date. However, given the small natural decay width of the Higgs boson, even small additional contributions from physics beyond the SM can lead to final states with substantial, and thus possibly detectable, branching fractions ($\mathcal{B}$)~\cite{Curtin:2013fra}. This Letter presents a search for Higgs boson decays into a $Z$~boson and a hadronically decaying light resonance in events with a same-flavor lepton pair (electrons or muons) and a jet in the ATLAS detector. Hadronic decays of an $\eta_c$ or of a $J/\psi$ charmonium resonance ($Q$), or of a light spin-0 boson from an extended Higgs sector with a mass up to 4~\GeV, are considered and are reconstructed as a single jet.
 
The Yukawa sector of the SM~\cite{Weinberg:1967tq} does not provide an explanation for the observed fermion mass hierarchy. As a result, a wide range of new physics scenarios have been proposed, including the Froggatt-Nielsen mechanism~\cite{FROGGATT1979277} and the Higgs-dependent Yukawa couplings model~\cite{Giudice:2008uua}; for a recent overview, see Ref.~\cite{deFlorian:2016spz}. The couplings of the Higgs boson to the third-generation fermions~\cite{HIGG-2018-04,CMS-HIG-18-016,HIGG-2018-13,CMS-HIG-17-035,HIGG-2017-07,CMS-HIG-16-043} have been observed, and a program to probe its couplings to the first- and second-generation charged leptons has been established~\cite{ATLAS-CONF-2019-028,Aad:2019ojw,Sirunyan:2018hbu,CMS-HIG-13-007}. For its couplings to first- and second-generation quarks, several approaches are being explored. Focusing on the Higgs boson's coupling to the charm quark, direct searches have been performed for Higgs boson decays into charm quarks~\cite{HIGG-2017-01,Sirunyan:2019qia} and for exclusive decays into a $J/\psi$ and a photon~\cite{HIGG-2016-23,CMS-HIG-14-003}, with no excess observed. Constraints from differential cross section measurements of Higgs boson production versus transverse momentum ($p_T$) have also been derived~\cite{ATLAS-CONF-2019-029,CMS-HIG-17-028}. Higgs boson decays into a gauge boson and a charmonium state, including an $\eta_c$ or a $J/\psi$, have been proposed as another way to access the coupling of the Higgs boson to the charm quark~\cite{Bodwin:2013gca,Kagan:2014ila,Chisholm:2016fzg}, and to probe the nature of the Higgs boson~\cite{Isidori:2013cla}. This search follows the last approach, and maximizes the signal acceptance by focusing on inclusive hadronic final states of the mesons in $H\to Z\eta_c$ and $H\to ZJ/\psi$ decays, which have SM branching fractions of $1.4\times 10^{-5}$ and $2.2\times 10^{-6}$~\cite{Isidori:2013cla}, respectively.
 
While the SM posits a single complex Higgs doublet field~\cite{Higgs:1964aa,Englert:1964aa}, extended Higgs sectors are motivated~\cite{Dawson:2018dcd} and provide a rich phenomenology of additional scalars. Two such models discussed here are the two-Higgs-doublet model (2HDM)~\cite{Curtin:2013fra,Delaunay:2013pja} and the 2HDM with an additional scalar singlet (2HDM$+$S)~\cite{Curtin:2013fra,Belyaev:2010ka}. These represent two of the simplest extensions of the scalar sector, and with their type-II fermion couplings they are necessary to generate the masses in the minimal supersymmetric SM and the next-to-minimal supersymmetric SM, respectively~\cite{Gunion:2002zf}. Both of these models can include additional light pseudoscalars ($a$) with significant $\mathcal{B}(H\to Za)$ or $\mathcal{B}(H\to aa)$~\cite{Curtin:2013fra}. In the 2HDM($+$S), these two $\mathcal{B}$ values can be adjusted independently, therefore searches for $H\to aa$ do not constrain $\mathcal{B}(H\to Za)$, so that searches for the latter decay are required~\cite{Curtin:2013fra,Chisholm:2016fzg}. Despite the Yukawa nature of the $a$ to fermion couplings, there are large regions of parameter space depending on the mass of $a$ and the ratio of the vacuum expectation values of the two Higgs-doublet fields ($\tan\beta$)~\cite{Curtin:2013fra}, where these pseudoscalars decay mainly to gluons and light up-type quarks, as the decays into down-type fermions are suppressed. These experimental signatures are also relevant in axion models~\cite{Bauer:2017ris,Kim:1986ax,Peccei:1977hh}, models of electroweak baryogenesis~\cite{Trodden:1998qg}, neutrino mass models~\cite{Camargo:2019ukv}, dark-matter models~\cite{Camargo:2019ukv,Arcadi:2018pfo}, and models of grand unification~\cite{Gunion:425736}. Previous searches for Higgs boson decays into light scalars have been performed at the Tevatron~\cite{Abazov:2009yi} and the LHC~\cite{CMS-HIG-17-029,CMS-HIG-16-015,CMS-HIG-14-019,CMS-HIG-13-010,HIGG-2014-02,HIGG-2016-01,EXOT-2016-22,HIGG-2016-03,HIGG-2017-09,EXOT-2013-24}. However, these were mostly focused on searches for $H\to aa$, in final states including leptons, photons or bottom quarks. By targeting the $H\to Za\text{,}\ a\to \text{hadrons}$ decay channel, this search accesses new, previously unexplored regions of the parameter space.
 
Searches for hadronic decays of light resonances are challenging at the LHC due to the large multijet background. However, substantial progress has been made in the use of jet substructure techniques in boosted final states~\cite{Asquith:2018igt}, typically in searches or measurements involving heavy resonances~\cite{Aad:2019uoz,Aad:2019wdr}. In this Letter, jet substructure variables enable the reconstruction of a light, boosted, hadronic final state. Information from the individual substructure variables is combined using machine learning techniques. Specifically, for event selection, a multilayer perceptron (MLP)~\cite{TMVA} classifier is employed. Given the range of masses considered, the classifier is provided with resonance-mass-related information from a separate MLP-based mass estimator, which results in improved classification performance over the full mass range.
 
This search is performed using the complete run~2 $pp$ collision dataset, produced between 2015 and 2018 at a center-of-mass energy $\sqrt{s}=13~\TeV$ by the LHC. The data were collected by the ATLAS detector~\cite{PERF-2007-01} and correspond to an integrated luminosity of $139~\text{fb}^{-1}$.
 
Monte Carlo (MC) samples of simulated events are used to model the signal selection efficiency. The signal samples were generated via the gluon-gluon fusion process using {\scriptsize POWHEG-BOX~v2}~\cite{Nason:2004rx,Frixione:2007vw,Alioli:2010xd}, with the {\scriptsize CT10} next-to-leading order (NLO) parton distribution function (PDF) set~\cite{Lai:2010vv}. Particle decays, hadronization, parton showers and the underlying event were modeled using {\scriptsize PYTHIA~v8.212}~\cite{Sjostrand:2007gs} and {\scriptsize Evt~Gen~v1.6.0}~\cite{Lange:2001uf}, interfaced to the {\scriptsize AZNLO}~\cite{Aad:2014xaa} set of tuned parameters and the {\scriptsize CTEQ6~L}1 PDF set~\cite{Pumplin:2002vw}. Next-to-next-to-leading order (NNLO) corrections are applied to the $p_T$ distribution of the Higgs boson. The $a$ branching fractions were determined using {\scriptsize PYTHIA~8}~\cite{Sjostrand:2007gs} with a 2HDM $\tan\beta$ value of 1, which predicts $a\to gg$ to be the dominant decay mode until $a\to c\bar c$ becomes kinematically accessible. The signal MC samples used in this analysis have $a$ masses of 0.5, 0.75, 1, 1.5, 2, 2.5, 3, 3.5, and 4~\GeV. The $Z$~boson is required to decay into pairs of electrons, muons, or $\tau$~leptons.
 
The background is dominated by $Z+\text{jets}$ events, modeled using {\scriptsize SHERPA~2.2.1}~\cite{Bothmann:2019yzt} interfaced to the {\scriptsize NNPDF~3.0} (NNLO) PDF set~\cite{Ball:2014uwa}. The inclusive production cross sections are known to NNLO in QCD~\cite{Anastasiou:2003ds}. The $ZZ$, $ZW$, and $t\bar t$ processes contribute $<1\%$ of the total background in this search. The diboson backgrounds were modeled using {\scriptsize SHERPA~2.2.1} interfaced to the {\scriptsize NNPDF~3.0} (NNLO) PDF set, except for gluon-induced $ZZ$ production, which was modeled using {\scriptsize SHERPA~2.2.2}~\cite{Bothmann:2019yzt}. All of the {\scriptsize SHERPA} samples used a set of tuned parameters developed by the {\scriptsize SHERPA} authors. The $t\bar t$ process was modeled using {\scriptsize POWHEG-BOX~v2}, while the subsequent decay, hadronization, parton shower and underlying event were modeled using {\scriptsize PYTHIA~v8.230} and {\scriptsize EvtGen~v1.6.0}. The {\scriptsize NNPDF~2.3} (LO) PDF set~\cite{Ball:2012cx} and the {\scriptsize A14} set of tuned parameters~\cite{ATL-PHYS-PUB-2014-021} were used.
 
The simulation of the ATLAS detector~\cite{SOFT-2010-01} in {\scriptsize GEANT4}~\cite{Agostinelli:2002hh} was used to model the interaction of particles with the detector in all the MC samples. Data-driven corrections are applied to the event-level trigger efficiencies, the jet vertex tagging efficiency~\cite{ATLAS-CONF-2014-018}, the electron~\cite{EGAM-2018-01} reconstruction, identification, and isolation efficiencies, and the muon~\cite{PERF-2015-10} reconstruction, isolation, and track-to-vertex association efficiencies.
 
Events are selected by a combination of single electron or muon triggers for each data-taking period~\cite{TRIG-2016-01,ATL-DAQ-PUB-2017-001,ATL-DAQ-PUB-2018-002,Aad:2019wsl}, and the online lepton reconstructed by the trigger is required to be within $\Delta R=0.1$~\cite{coordinate} of an off-line reconstructed lepton. Events are required to have at least one reconstructed primary interaction vertex~\cite{ATL-PHYS-PUB-2015-026}. Electron candidates are reconstructed by matching tracks in the inner detector to topological energy clusters in the electromagnetic calorimeter~\cite{EGAM-2018-01}, and must pass a likelihood-based selection which requires the shower profile to be compatible with that of an electromagnetic shower. Muons are reconstructed using tracks in the muon spectrometer, matched to tracks in the inner detector where available~\cite{PERF-2014-05}. Electrons and muons are each required to have $p_T>18~\GeV$, and at least one must have $p_T>27~\GeV$. Electrons (muons) are required to be reconstructed within $\vert\eta\vert<2.47$ ($\vert\eta\vert<2.7$), but electrons within $1.37<\vert\eta\vert<1.52$ are excluded. The transverse energy sum in a cone of size $\Delta R=0.2$ around the electron [muon] in the calorimeter must be less than $20\%$ ($30\%$) of the lepton's $p_T$, and the summed $p_T$ of tracks within a cone of variable size $\Delta R=\text{min}(0.2,~10~\GeV/p_T)$ [$\Delta R=\text{min}(0.15,~10~\GeV/p_T)$] around the electron [muon] must be less than $15\%$ of its $p_T$. Contributions from nearby electrons and muons are removed from these cones. If an inner detector track is present, muons must also have a longitudinal impact parameter $\vert z_0\sin\theta\vert < 0.5$~mm, and a transverse impact parameter $\vert d_0\vert < 1$~mm, relative to the primary interaction vertex. At least two same-flavor opposite-sign electrons or muons are required to pass this selection, and have an invariant mass compatible with the mass of the $Z$~boson: $81<m_{\ell\ell}<101~\GeV$. If multiple same-flavor opposite-sign lepton pairs fulfill this requirement, the pairing with an invariant mass closest to that of the $Z$~boson is chosen. $Z\to\tau\tau$ decays are reconstructed through the leptonic decays of the $\tau$~leptons.
 
The hadronically decaying resonance is reconstructed as a single jet using the anti-$k_t$ jet algorithm~\cite{Cacciari:2008gp,Fastjet} with a radius parameter of 0.4, formed from topological calorimeter energy clusters~\cite{PERF-2014-07,ATL-PHYS-PUB-2015-036} and calibrated to the electromagnetic energy scale. Jet energies are corrected for contributions from simultaneous inelastic $pp$ interactions (pileup) using a jet-area-based technique~\cite{Cacciari:2007fd,Aad:2015ina}, and calibrated~\cite{PERF-2011-03,PERF-2016-04} using $p_T$- and $\eta$-dependent correction factors determined from simulation, with residual corrections from \emph{in situ} measurements applied to data and internal jet properties. Jets are required to have $p_T>20~\GeV$ and $\vert\eta\vert<2.5$, and satisfy a jet cleaning requirement~\cite{ATLAS-CONF-2015-029}. To reject jets from pileup interactions, jets with $p_T<60~\GeV$ and $\vert\eta\vert<2.4$ are required to pass a ``jet vertex tagger''~\cite{ATLAS-CONF-2014-018} requirement. An overlap removal procedure resolves cases in which multiple electrons, muons or jets are reconstructed from the same detector signature. Higgs boson candidates are reconstructed from the lepton pair and jet system, which is required to have an invariant mass passing a loose preselection requirement: $m_{\ell\ell j}<250~\GeV$. If multiple jets satisfy these requirements, the jet with the highest $p_T$ is selected. The acceptance for this preselection, evaluated using generator-level MC samples, varies between 28\% and 29\% for the different $Q/a$ signal hypotheses.
 
\begin{figure}[]
\centering
\subfigure[]{\includegraphics[width=0.495\textwidth]{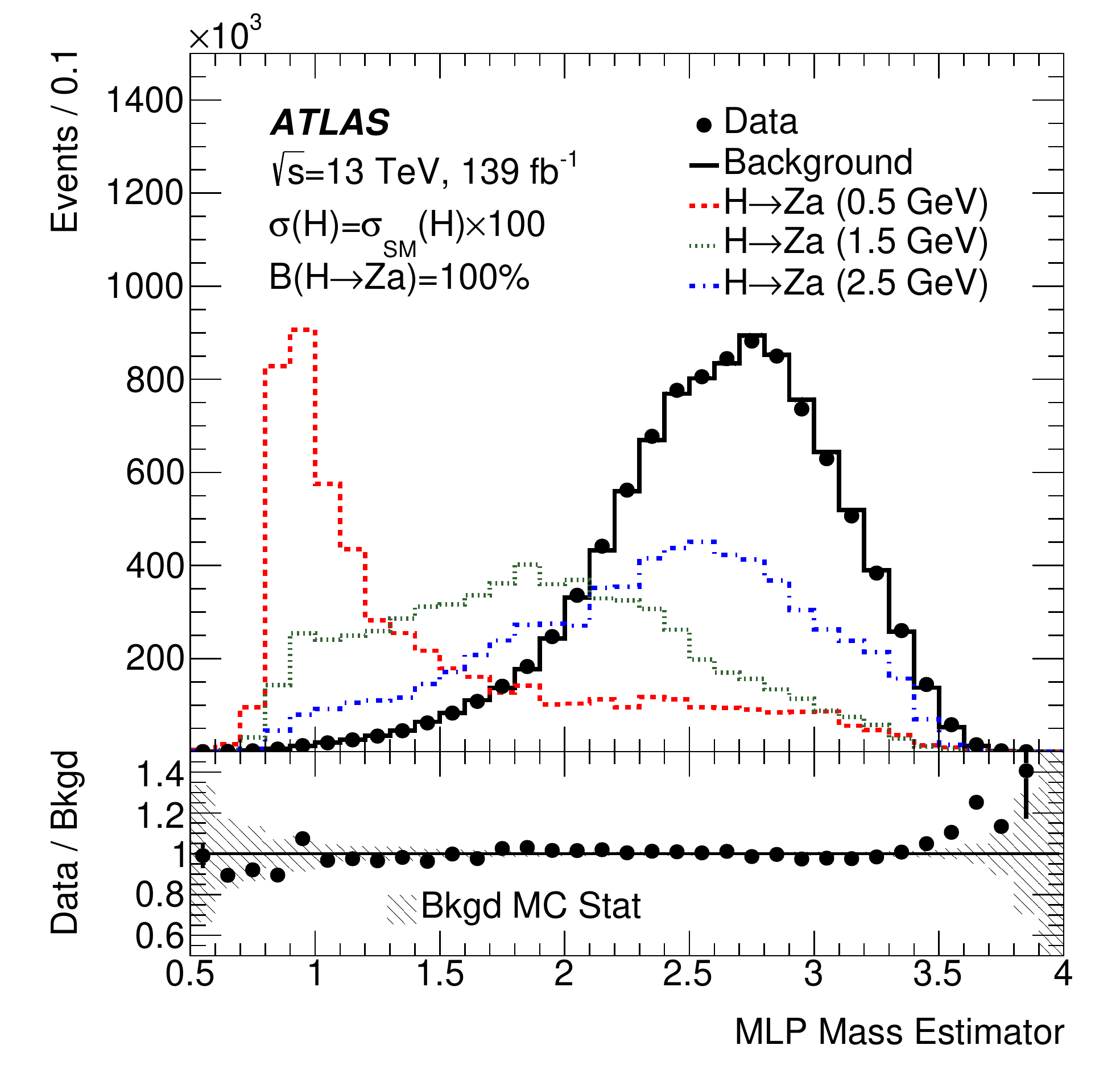}}
\subfigure[]{\includegraphics[width=0.495\textwidth]{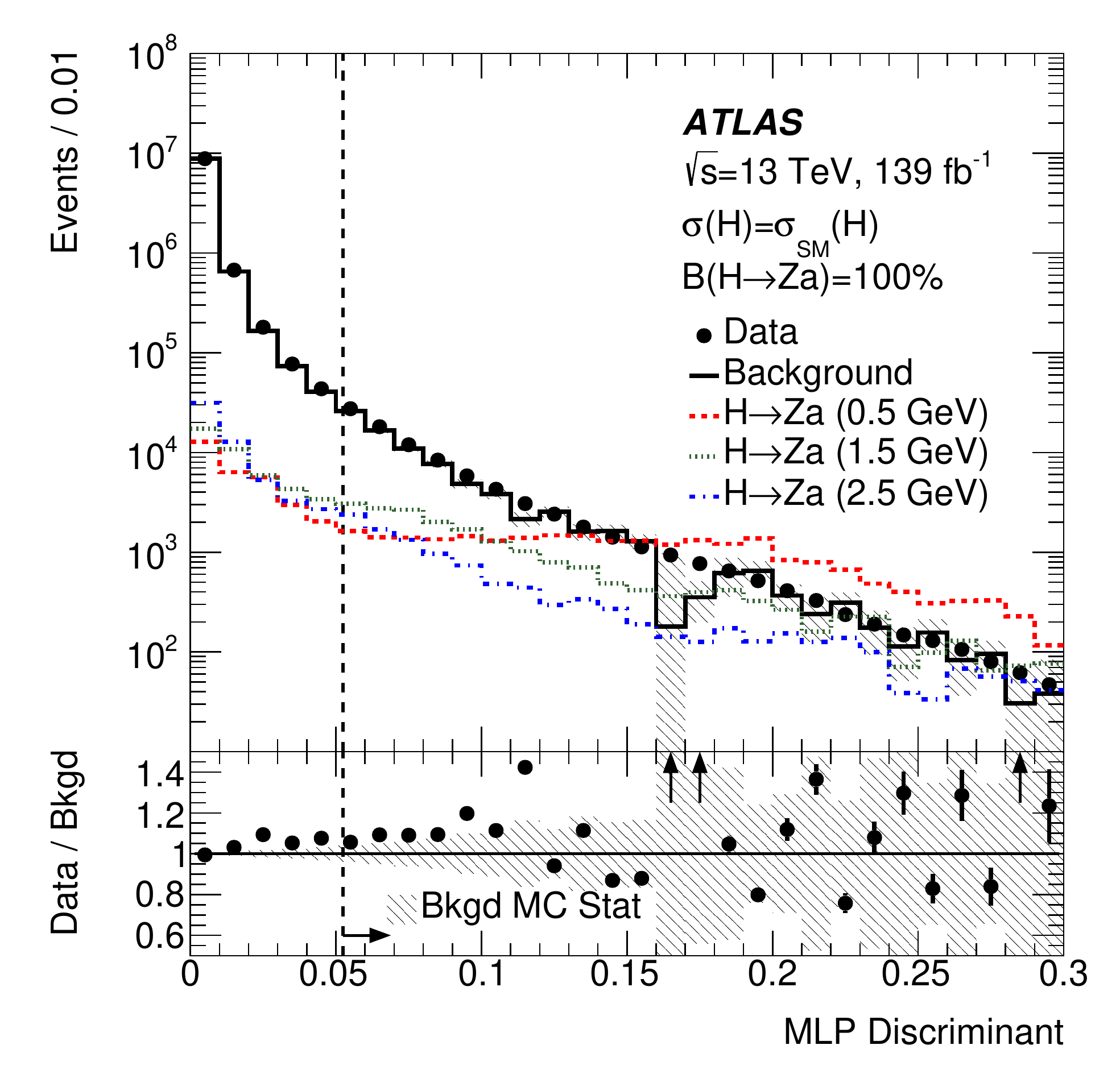}}
\caption{Output of (a) the regression and (b) the classification MLPs, for data, background, and three signal hypotheses. Events are required to pass the complete event selection, including the $120<m_{\ell\ell j}<135~\GeV$ requirement, but not the requirement on the classification MLP output variable. The background normalization is set equal to that of the data, and the signal normalizations assume the SM Higgs boson inclusive production cross section and $\mathcal{B}(H\to Za)=100\%$, and in (a) the signal normalization is scaled up by a factor of 100. The error bars (hatched regions) represent the data (MC) sample statistical uncertainty, in both the histograms and the ratio plots. In (b) the region to the right of the dashed line is the signal region.}
\label{fig:mlps}
\end{figure}

MLPs~\cite{TMVA} are used to select signal events passing this preselection. The MLP input variables are built using tracks matched to the calorimeter jet by ghost association~\cite{Cacciari:2007fd}, in which the tracks are included in the jet clustering process as with negligible energy and their angles from the jet axis. This allows the MLP to benefit from the high resolution of the tracking detector. These tracks must have $p_T>500~\MeV$ and $\vert\eta\vert<2.5$, and pass loose quality and track-to-vertex association requirements~\cite{PERF-2015-08} to reject fake tracks from the reconstruction and tracks from pileup, respectively. Six dimensionless variables are constructed using these tracks: the ratio of the $p_T$ of the highest $p_T$ track to the $p_T$ of the ghost-associated track system; the angular separation $\Delta R$ between the highest $p_T$ track and the calorimeter jet axis; NSubJettiness~2~\cite{Thaler:2010tr}, using exclusive-$k_t$ subjet axes with radius parameters of 0.2, and a jet axis radius parameter of 0.4; $\text{angularity}(2)$~\cite{Almeida:2008yp}; and $U_1(0.7)$ and $M_2(0.3)$, which are modified energy correlation functions~\cite{Moult:2016cvt} designed for quark-gluon discrimination and to target two-pronged substructure, respectively. These variables primarily capitalize on the presence of a narrow resonance or two-pronged substructure in the track system. Initially, a \emph{regression} MLP~\cite{TMVA}, using four hidden layers of 12 nodes, is trained using the above input variables and the $a$ signal samples to estimate the mass of $a$, as shown in Fig.~\ref{fig:mlps}(a). This estimated mass is then provided alongside the six input variables to a \emph{classification} MLP~\cite{TMVA}, to inform the classifier about the part of the hadronic resonance mass spectrum where the specific event lies. This classification MLP has two hidden layers of six and five nodes and is trained using the $a$ signal samples and the background samples. The 0.75~\GeV{} $a$ signal sample is excluded from the training of the classification MLP to ensure an even spacing between the $a$ mass hypotheses, so the training is not biased toward lower masses. Both MLPs use sigmoidal response functions with summed inputs and are trained using backpropagation with a mean-square estimator~\cite{TMVA}, as these resulted in optimal discrimination without overtraining. The addition of the regression MLP was found to result in about a $13\%$ improvement in the $S/\sqrt{B}$ of the classification MLP, where $S$ and $B$ are the expected numbers of signal and background events passing the MLP requirement, respectively. The classification MLP output variable ($M$) is shown in Fig.~\ref{fig:mlps}(b).
 
The signal region (SR) for this search is defined by the requirements $120<m_{\ell\ell j}<135~\GeV$ and $M>0.0524$, chosen to maximize the expected $S/\sqrt{B}$, averaged over the various $a$ mass hypotheses. The efficiency of this MLP requirement for events passing the preselection is $(0.761\pm 0.020)\%$ for the background, $(5.89 \pm 0.24)\%$ and $(6.66 \pm 0.26)\%$ for $H\to Z\eta_c$ and $H\to ZJ/\psi$, respectively, and between $(1.88 \pm 0.15)\%$ and $(45.9 \pm 0.8)\%$ for $H\to Za$. The efficiencies for the complete selection are estimated using MC samples and are $(0.545 \pm 0.022)\%$ and $(0.560 \pm 0.022)\%$ for $H\to Z\eta_c$ and $H\to ZJ/\psi$, respectively, and range between $(0.140 \pm 0.011)\%$ and $(3.27 \pm 0.06)\%$ for $H\to Za$. The efficiencies are highest for the lowest $a$ mass hypotheses, due to higher probabilities to pass the MLP requirement. The efficiency for $H\to Z\eta_c$ events to pass the MLP requirement is lower than that of $H\to ZJ/\psi$ events, as $J/\psi$ decays tend to have a lower charged hadron multiplicity. Using the predicted cross section for inclusive SM Higgs boson production of $55.7^{+3.0}_{-3.9}~\text{pb}$~\cite{Dittmaier:2011ti}, and $\mathcal{B}[H\to Z(Q/a)]=100\%$, gives expected signal yields of $4260$ and $4370$ for $H\to Z\eta_c$ and $H\to ZJ/\psi$, respectively, and between $1090$ and $25~600$ for $H\to Za$.
 
A ``modified ABCD estimate'' of the total background in the SR is derived using four regions: $A$, defined by $0.0341<M<0.0524$, expected to contain about $10\%$ of the total background, and $155~\GeV < m_{\ell\ell j} < 175~\GeV$; $B$, defined by the $m_{\ell\ell j}$ requirement of the SR and the $M$ requirement of region $A$; $C$, defined by the $M$ requirement of the SR and the $m_{\ell\ell j}$ requirement of $A$; and $D$, which is the SR. An initial data-driven background estimate in the SR is calculated as $D=BC/A$, then MC samples, reweighted to match data, are used to correct this estimate for the 13\% correlation between the $m_{\ell\ell j}$ and $M$ variables. This reweighting is performed in the $p_T$ of the calorimeter jet, the number of ghost-associated tracks and $U_1(0.7)$. This background estimate is $82~400\pm 2900$ events in the SR, where the uncertainty is due to the limited data and MC sample statistics. The background estimation method is found to be consistent with data within $1.7$ times the total statistical and systematic uncertainty in 14 validation regions, defined in regions of the $m_{\ell\ell j}$ and $M$ variables.
 
A measure of $\sigma(pp\to H)\mathcal{B}[H\to Z(Q/a)]$ is extracted for a given signal hypothesis using a maximum-likelihood fit~\cite{Cowan:2010js-witherratum} to the number of events observed in the SR. The systematic uncertainties are included in the likelihood fit as nuisance parameters which modify the signal efficiencies or the simulation-based correction used to calculate the expected background yield. These systematic uncertainties include uncertainties in the signal and background modeling, and experimental uncertainties. The sources of modeling uncertainty include the limited MC sample statistics, renormalization scale and choice of MC generator for the signal and background, and a signal uncertainty to account for the extrapolation from gluon-gluon fusion signal samples to the inclusive Higgs boson production cross section. The effects of factorization scale and PDF uncertainties are found to be negligible. The experimental uncertainties considered are due to the luminosity~\cite{ATLAS-CONF-2019-021}, pileup~\cite{Aad:2019cpw}, triggers, lepton~\cite{PERF-2017-03,PERF-2017-01,PERF-2015-10} and jet~\cite{PERF-2016-04} reconstruction. The total uncertainty on the extracted signal yield is dominated by the background modeling uncertainties, the largest being due to limited MC sample statistics. The total uncertainty on the background in the SR is $3700$ events, where the uncertainty due to the limited data and MC sample statistics is $2900$ and the modeling uncertainty is $2300$. The data statistical uncertainty corresponds to approximately $8\%$ of the total uncertainty on the extracted signal yield.
 
\begin{figure}[!htbp]
\centering
\includegraphics[width=0.495\textwidth]{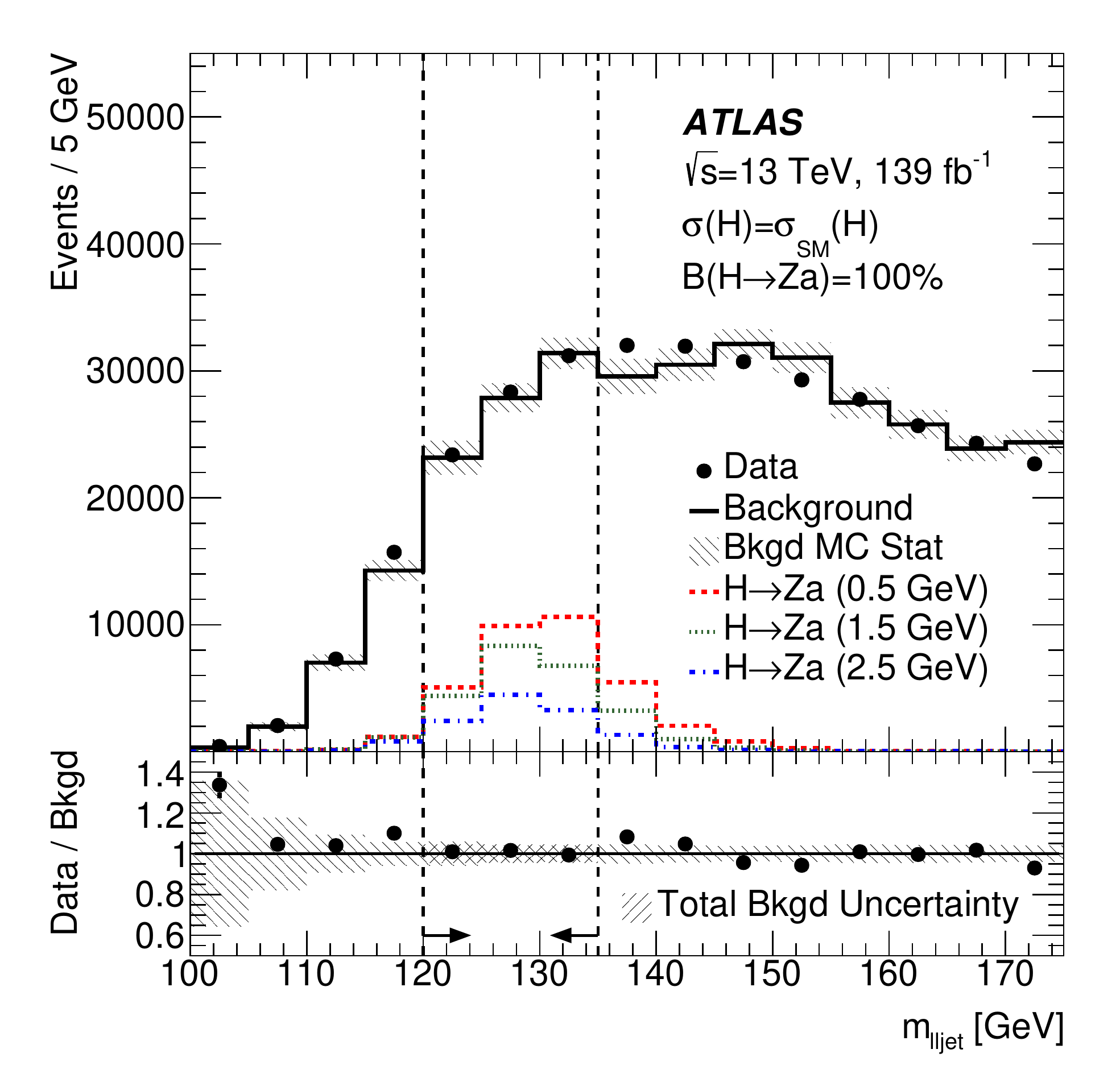}
\caption{Invariant mass of the lepton pair plus jet system, for data, background and three signal hypotheses. Events are required to pass the complete event selection, including the MLP output variable requirement, but not the $120<m_{\ell\ell j}<135~\GeV$ requirement. The background normalization is defined by the background estimate in the signal region, and the signal normalizations assume the SM Higgs boson inclusive production cross section and $\mathcal{B}(H\to Za)=100\%$. The error bars (hatched regions) represent the data (MC) sample statistical uncertainty, in both the histograms and the ratio plots. The region between the vertical dashed lines is the signal region. The total background uncertainty in the signal region is also indicated.}
\label{fig:mllj}
\end{figure}

The SR contains $82~908$ data events. This result is compatible with the SM background-only expectation, and the three-body mass distribution is shown in Fig.~\ref{fig:mllj}. Upper limits at 95\% confidence level (CL) are set on $\sigma(pp\to H)\mathcal{B}[H\to Z(Q/a)]$ for the various signal hypotheses, using the profile-likelihood test statistic~\cite{Cowan:2010js-witherratum} and the $\text{CL}_\text{s}$ technique~\cite{Read:2002hq}. The observed (expected) upper limits for the $H\to Z\eta_c$ and $H\to ZJ/\psi$ hypotheses are $110$ and $100~\text{pb}$ ($100^{+40}_{-30}$ and $100^{+40}_{-30}~\text{pb}$), respectively, while the upper limits for the $H\to Za$ signal hypotheses are given in Table~\ref{tab:observedmodelindepedentresults}. In the absence of systematic uncertainties, these limits would range between $1.9$ and $55~\text{pb}$ for the different signal hypotheses. To simplify the interpretation, the upper limits are quoted for $\mathcal{B}(a\to gg)=100\%$ and $\mathcal{B}(a\to s\bar s)=100\%$. Because of the Yukawa ordering of the decays of Higgs bosons, only decays into gluon and strange quark pairs are considered. The tighter limits for the $a \to s\bar s$ decays are due to a higher MLP selection efficiency. The systematic uncertainties for $a \to gg$ and $a \to s\bar s$ decay hypotheses are estimated using the inclusive decays as modeled in {\scriptsize PYTHIA~8}, which is a good approximation due to the dominance of the background modeling uncertainties. This assumption allows a limit to be set on the decay of $a$ into $gg$ or $s\bar s$ final states, in any ratio, by using a weighted sum of the two limits. Higgs boson decays to a $Z$~boson and a quarkonium state other than the considered signal process are not included in the statistical interpretation.
 
\begin{table}[!htbp]{\footnotesize\renewcommand{\arraystretch}{1.25}
\begin{center}
\footnotesize
\caption{Expected and observed 95\% $\text{CL}$ upper limits on $\sigma(pp\to H)\mathcal{B}(H\to Za)/\text{pb}$. These results are quoted for $\mathcal{B}(a\to gg)=100\%$ and $\mathcal{B}(a\to s\bar s)=100\%$ for each signal sample.}
\vspace{1mm}
\begin{tabular}{ccrcc}
\toprule
\multirow{2}{*}{$a$ mass [\GeV]} & \multicolumn{2}{c}{$a\to gg$} & \multicolumn{2}{c}{$a \to s\bar s$} \\
& Exp & Obs & Exp & Obs \\
\midrule
0.5 & $16^{+6}_{-5}$ & 17 & & \\
~~0.75 & $19^{+7}_{-5}$ & 20 & & \\
1.0 & $17^{+7}_{-5}$ & 18 & & \\
1.5 & $20^{+8}_{-6}$ & 22 & $19^{+7}_{-5}$ & 20 \\
2.0 & $~26^{+10}_{-7}$ & 27 & $23^{+9}_{-6}$ & 24 \\
2.5 & $~38^{+15}_{-11}$ & 40 & $~32^{+12}_{-9}$ & 33 \\
3.0 & $~75^{+29}_{-21}$ & 78 & $~65^{+25}_{-18}$ & 68 \\
3.5 & $110^{+40}_{-30}~$ & 120 & & \\
4.0 & $\,320^{+130}_{-90}$ & 340 & & \\
\bottomrule
\end{tabular}
\label{tab:observedmodelindepedentresults}
\end{center}}
\end{table}
 
In conclusion, a search has been performed for Higgs boson decays into a $Z$~boson and either a $\eta_c$ or $J/\psi$ charmonium state, or a light spin-0 boson. No excess is found, and 95\% $\text{CL}$ upper limits are set on $\sigma(pp\to H)\mathcal{B}[H\to Z(Q/a)]$, with values of $110$ and $100~\text{pb}$ for the $H\to Z\eta_c$ and $H\to ZJ/\psi$ hypotheses, respectively, and with values in the range $17\text{--}340~\text{pb}$ for the $H\to Za$ signal hypotheses. Assuming the SM prediction for inclusive Higgs boson production, the limits on charmonium decay modes correspond to branching fraction limits in excess of 100\%. This is the first direct limit on decays of the observed Higgs boson to light scalars, decaying to light quarks or gluons. Because of the large value of $\mathcal{B}(a \to \text{hadrons})$ over the entire 2HDM($+$S) parameter space, these limits represent tight, direct constraints for low (high) $\tan\beta$ in the type-II and type-III (type-VI) 2HDM$+$S~\cite{ATL-PHYS-PUB-2018-045}.


We thank CERN for the very successful operation of the LHC, as well as the
support staff from our institutions without whom ATLAS could not be
operated efficiently.
 
We acknowledge the support of ANPCyT, Argentina; YerPhI, Armenia; ARC, Australia; BMWFW and FWF, Austria; ANAS, Azerbaijan; SSTC, Belarus; CNPq and FAPESP, Brazil; NSERC, NRC and CFI, Canada; CERN; ANID, Chile; CAS, MOST and NSFC, China; COLCIENCIAS, Colombia; MSMT CR, MPO CR and VSC CR, Czech Republic; DNRF and DNSRC, Denmark; IN2P3-CNRS and CEA-DRF/IRFU, France; SRNSFG, Georgia; BMBF, HGF and MPG, Germany; GSRT, Greece; RGC and Hong Kong SAR, China; ISF and Benoziyo Center, Israel; INFN, Italy; MEXT and JSPS, Japan; CNRST, Morocco; NWO, Netherlands; RCN, Norway; MNiSW and NCN, Poland; FCT, Portugal; MNE/IFA, Romania; JINR; MES of Russia and NRC KI, Russian Federation; MESTD, Serbia; MSSR, Slovakia; ARRS and MIZ\v{S}, Slovenia; DST/NRF, South Africa; MICINN, Spain; SRC and Wallenberg Foundation, Sweden; SERI, SNSF and Cantons of Bern and Geneva, Switzerland; MOST, Taiwan; TAEK, Turkey; STFC, United Kingdom; DOE and NSF, United States of America. In addition, individual groups and members have received support from BCKDF, CANARIE, Compute Canada, CRC and IVADO, Canada; Beijing Municipal Science \& Technology Commission, China; COST, ERC, ERDF, Horizon 2020 and Marie Sk{\l}odowska-Curie Actions, European Union; Investissements d'Avenir Labex, Investissements d'Avenir Idex and ANR, France; DFG and AvH Foundation, Germany; Herakleitos, Thales and Aristeia programmes co-financed by EU-ESF and the Greek NSRF, Greece; BSF-NSF and GIF, Israel; La Caixa Banking Foundation, CERCA Programme Generalitat de Catalunya and PROMETEO and GenT Programmes Generalitat Valenciana, Spain; G\"{o}ran Gustafssons Stiftelse, Sweden; The Royal Society and Leverhulme Trust, United Kingdom.
 
The crucial computing support from all WLCG partners is acknowledged gratefully, in particular from CERN, the ATLAS Tier-1 facilities at TRIUMF (Canada), NDGF (Denmark, Norway, Sweden), CC-IN2P3 (France), KIT/GridKA (Germany), INFN-CNAF (Italy), NL-T1 (Netherlands), PIC (Spain), ASGC (Taiwan), RAL (UK) and BNL (USA), the Tier-2 facilities worldwide and large non-WLCG resource providers. Major contributors of computing resources are listed in Ref.~\cite{ATL-SOFT-PUB-2020-001}.
 
 
\printbibliography
 
\clearpage
 
\begin{flushleft}
\hypersetup{urlcolor=black}
{\Large The ATLAS Collaboration}

\bigskip

\AtlasOrcid[0000-0002-6665-4934]{G.~Aad}$^\textrm{\scriptsize 102}$,    
\AtlasOrcid[0000-0002-5888-2734]{B.~Abbott}$^\textrm{\scriptsize 128}$,    
\AtlasOrcid[0000-0002-7248-3203]{D.C.~Abbott}$^\textrm{\scriptsize 103}$,    
\AtlasOrcid[0000-0002-2788-3822]{A.~Abed~Abud}$^\textrm{\scriptsize 36}$,    
\AtlasOrcid[0000-0002-1002-1652]{K.~Abeling}$^\textrm{\scriptsize 53}$,    
\AtlasOrcid[0000-0002-2987-4006]{D.K.~Abhayasinghe}$^\textrm{\scriptsize 94}$,    
\AtlasOrcid[0000-0002-8496-9294]{S.H.~Abidi}$^\textrm{\scriptsize 167}$,    
\AtlasOrcid[0000-0002-8279-9324]{O.S.~AbouZeid}$^\textrm{\scriptsize 40}$,    
\AtlasOrcid{N.L.~Abraham}$^\textrm{\scriptsize 156}$,    
\AtlasOrcid[0000-0001-5329-6640]{H.~Abramowicz}$^\textrm{\scriptsize 161}$
% ...
\bigskip
\\

% ...
$^{36}$CERN, Geneva; Switzerland.\\
% ...
$^{40}$Niels Bohr Institute, University of Copenhagen, Copenhagen; Denmark.\\
% ...
$^{53}$II. Physikalisches Institut, Georg-August-Universit\"{a}t G\"ottingen, G\"ottingen; Germany.\\
% ...
$^{94}$Department of Physics, Royal Holloway University of London, Egham; United Kingdom.\\
% ...
$^{102}$CPPM, Aix-Marseille Universit\'e, CNRS/IN2P3, Marseille; France.\\
$^{103}$Department of Physics, University of Massachusetts, Amherst MA; United States of America.\\
% ...
$^{128}$Homer L. Dodge Department of Physics and Astronomy, University of Oklahoma, Norman OK; United States of America.\\
% ...
$^{156}$Department of Physics and Astronomy, University of Sussex, Brighton; United Kingdom.\\
% ...
$^{161}$Raymond and Beverly Sackler School of Physics and Astronomy, Tel Aviv University, Tel Aviv; Israel.\\
% ...
$^{167}$Department of Physics, University of Toronto, Toronto ON; Canada.\\
% ...

\end{flushleft}

 
\end{document}